# COLLECTIVE MODE MINING FROM MOLECULAR DYNAMICS SIMULATIONS: A COMPARATIVE APPROACH

VITO DARIO CAMIOLA

*Istituto Nanoscienze del Cnr and NEST-Scuola Normale Superiore, Piazza San Silvestro 12*
*Pisa, 56127, Italy*
*dario.camiola@gmail.com*

VALENTINA TOZZINI[*]

*Istituto Nanoscienze del Cnr and NEST-Scuola Normale Superiore, Piazza San Silvestro 12*
*Pisa, 56127, Italy*
*valentina.tozzini@nano.cnr.it*
*http://www.nano.cnr.it/index.php?mod=peo&id=256*

The evaluation of collective modes is fundamental in the analysis of molecular dynamics simulations. Several methods are available to extract that information, i.e normal mode analysis, principal component and spectral analysis of trajectories, basically differing by the quantity considered as the nodal one (frequency, amplitude, or pattern of displacement) and leading to the definition of different kinds of collective excitations and physical spectral observables. Different views converge in the harmonic regime and/or for homo-atomic systems. However, for anharmonic and out of equilibrium dynamics different quantities bring different information and only their comparison can give a complete view of the system behavior. To allow such a comparative analysis, we review and compare the different approaches, applying them in different combination to two examples of physical relevance: graphene and fullerene $C_{60}$.

*Keywords*: Vibrational modes; spectral analysis; principal modes analysis; molecular dynamics simulations; graphene; fullerene.

## 1. Introduction

Collective dynamics analysis is a fundamental strategy to analyze the behavior of molecular systems of any nature and complexity, and a direct way to compare theories with information coming from a number of experimental techniques such as IR and Raman spectroscopy [Sathyabnaryana (2004)]. The theory underlying the vibrational modes definition has been established long ago, at least in the harmonic regime [Cui and Bahar (2006)], (Born and Huang (1962)] and basically relies on the evaluation and of the dynamical matrix **[**Zimmermann *et al* (2011)]**,** [Baroni et al (2001)] describing the forces acting on a system close to an equilibrium configuration. However, this approach is rather restrictive: being purely mechanical (not dynamical) it gives no information on the population of modes, or on their interactions due to anharmonicity. The latter can be included within a perturbative approach, which, however, becomes increasingly complex as the system get farther from equilibrium configuration.

---

[*] Corresponding author. Phone: +39 050 509433. Fax +39 050 509417

Conversely, approaches based on the post-processing of trajectories from molecular dynamics (MD) simulations [Koukaras *et al* (2015)] naturally include anharmonicity to any order with no additional computational cost and independently on the Hamiltonian complexity; they account for the relative population of the modes in different thermal and environmental conditions (also possibly more similar to the experimental ones).

There are different methods to extract collective motions from a MD simulation, focusing different aspect of the mode. Specifically, the spectral analysis of the time correlation functions, primarily returns information on frequencies and mode populations [Kohanoff (1994)], while the principal component analysis of displacements focuses on the geometrical and space coherency aspects of the collective motion [Amadei (1993)]. Therefore a comparative analysis can give a more accurate representation and interpretation of the system dynamics in the anharmonic regime.

In this work, we comparatively review those methods, which were implemented in a home-made software (available upon request). We remark differences and complementarity and highlight the applicability to different situations.

## 2. An overview on collective modes analysis: basic theory

Collective vibrational excitations are dynamical states ("modes") of a system in which atoms move coherently, generally identified by specific frequencies and patterns of atomic displacement. The intensity of the excitation is classically related to the squared amplitude of oscillation, which is superseded by the mode population concept in quantum mechanics. These established ideas are easily derived within the theory of harmonic systems[Cui and Bahar (2006)], [Born and Huang (1962)], which also gives operative ways to evaluate the mentioned observables. However, when anharmonicity appears, and the complexity of the system increases, those concepts become less well defined. We review in this section concepts and definitions in order to establish a common formal framework.

### 2.1. *Normal modes analysis*

We first review the definition of normal modes in the harmonic regime. It is convenient to define both displacement and mass weighted displacement vectors as follows

$$\delta \mathbf{r} = \mathbf{r} - \mathbf{r}_0 = (\delta \mathbf{r}_1, \cdots, \delta \mathbf{r}_n) \qquad \mathbf{u} = (\delta \mathbf{r}_1 \sqrt{m_1}, \cdots, \delta \mathbf{r}_n \sqrt{m_n}) \qquad (1)$$

subscript 0 indicating the equilibrium coordinate and *n* being the atom index. When not explicitly stated, we will adopt the following conventions: not-indexed-boldface font for 3N vectors (and operators), with both atomic and Cartesian indexes implicitly implied; indexed-boldface font for Cartesian 3D vectors (or operators) relative to single atoms quantities (in this case only Cartesian index is implied), italics for scalars or vector/matrix elements. Whenever not explicitly indicated, the Cartesian and atomic indexes are jointed (e.g. *i* and *j* in the following eqns range both on the atoms and Cartesian coordinates).

The physical quantity involved in the definition of normal modes (NM) is the dynamical matrix, related to the Hessian of the potential energy as follows

$$D_{ij} = \left.\frac{\partial^2 U}{\partial u_i \partial u_j}\right|_0 = \left.\frac{\partial^2 U}{\partial r_i \partial r_j}\right|_0 \frac{1}{\sqrt{m_i m_j}} = \frac{K_{ij}}{\sqrt{m_i m_j}} \qquad \mathbf{D} = \mathbf{M}^{-1/2}\mathbf{K}\mathbf{M}^{-1/2} \qquad M^{1/2}_{ij} = \delta_{ij}\sqrt{m_i} \qquad (2)$$

(the operator form is used in the central equation). The squared root mass operator $\mathbf{M}^{1/2}$ is diagonal in the Cartesian space, as its inverse $\mathbf{M}^{-1/2}=(\mathbf{M}^{1/2})^{-1}$. Last eqn (1), can also be written as $\mathbf{u} = \mathbf{M}^{1/2}\delta\mathbf{r}$. NMs are defined through the standard eigenvalues problem $\mathbf{Du}=\lambda\mathbf{u}$, whose solution exactly decouples the equation of motion of a harmonic system in a set of simple oscillators. As a consequence

$$\mathbf{u}(t) = \sum_I \mathbf{e}_I \Re[Q_I^0 \exp(i\omega_I t)] \qquad \delta\mathbf{r}(t) = \sum_I \mathbf{M}^{-1/2}\mathbf{e}_I \Re[Q_I^0 \exp(i\omega_I t)] \qquad \mathbf{e}_I = (a_{I,1},\cdots,a_{I,n}) \qquad (3)$$

$\mathbf{e}_I$ being the normalized eigenvector (3N-long, N being the number of atoms) of the I-th mode, $Q_I$ its amplitude and $\omega_I=\sqrt{\lambda_I}$ its frequency. Since $\mathbf{D}$ is symmetric and real, $\{\mathbf{e}_I\}$ can be chosen as an orthonormal basis, and the transformation matrix $a_{Ij}$ between this and the Cartesian basis is unitary. We remark that $\mathbf{M}^{-1/2}$ is not diagonal on the $\{\mathbf{e}_I\}$ basis, which makes the dependence of the Cartesian displacement $\delta\mathbf{r}$ on the masses rather complex, in general. The only simple case is the homo-atomic system (all equal masses), in which case $\mathbf{M}^{1/2}=m^{1/2}\mathbf{1}$ and $\mathbf{M}^{-1/2}=m^{-1/2}\mathbf{1}$ ($\mathbf{1}$ being the identity operator).

NM analysis (NMA) only includes *mechanical information* about the (harmonic part of the) Hamiltonian, namely, no dynamic or thermodynamic information about the modes population (amplitude) can be extracted from it. If a dynamical trajectory is available for the system, however, the (time dependent) amplitude of a given mode can be obtained projecting it onto the I-th eigenvector:

$$\mathbf{u}_I(t) = (\mathbf{u}(t)\cdot\mathbf{e}_I)\mathbf{e}_I = Q_I(t)\mathbf{e}_I \qquad \delta\mathbf{r}_I(t) = \mathbf{M}^{-1/2}\mathbf{u}_I(t) = Q_I(t)\mathbf{M}^{-1/2}\mathbf{e}_I = Q_I(t)\boldsymbol{\xi}_I \qquad (4)$$

with $Q_I(t)=\mathbf{u}(t)\cdot\mathbf{e}_I$ being a simple oscillation $Q_I(t)=\Re[Q_I^0 \exp(i\omega_I t)]$ in the purely harmonic case. In general, however, $Q_I(t)$ might include exponential dumping, multiple frequency or dispersion. Therefore eqn (4) and deviation of $Q_I(t)$ from a pure oscillation can be used to estimate anharmonicity. The supporting vector basis $\boldsymbol{\xi}_I$ is defined to express the non-mass weighted modes coordinates as a function of the mass weighted ones. However, we remark that the projection $\delta\mathbf{r}(t)\cdot\mathbf{e}_I$ of the non-mass weighted trajectory *does not* return a single frequency time dependent amplitude even in the purely harmonic case, due to the general non-orthogonality of the vectors $\boldsymbol{\xi}_I=\mathbf{M}^{-1/2}\mathbf{e}_I$ to $\mathbf{e}_J$, except in the obvious case of homo-atomic system, where $\delta\mathbf{r}(t)=\mathbf{u}(t)/\sqrt{m}$ and $\boldsymbol{\xi}_I=\mathbf{e}_I/\sqrt{m}$.

To complete this scheme, we observe that, in crystals, the additional symmetries by translation of lattice vectors imply that the mode index $I$ can be split into a branch index $s$ and a "wave vector" $k$ belonging to the Brillouin zone (BZ) of the reciprocal lattice and $\omega_I \to \omega_s(\mathbf{k})$ is called the dispersion relation of the branch $s$ [Ashcroft and Mermin (1976)]. $\mathbf{k}$ is in principle continuous, reflecting the infinite spatial extension of a crystal, but in practice it is sampled on a discrete set values. Using any finite $\mathbf{k}$ within the BZ is equivalent to describing a wave with wavelength larger than a single unit cell, or, in other words, to use Born-von-Karman periodic conditions involving a number of unit cells commensurate with that wavelength. This has relevance especially if normal modes have to be detected by mean of a simulation.

**2.2. Principal modes analysis**

While NMA extracts frequencies and mode patterns from the forces, the analysis of principal components (PCA) of the displacement fluctuations, also called principal modes analysis, takes a different point of view, using as input a trajectory and focusing on the amplitude and direction of the displacements. The method is based on the standard eigenvalues problem resolution for the covariance matrix **C** [Amadei (1993)]. The corresponding eigenvectors are named principal modes. As for NMA, it is convenient to define both the non-mass weighted and mass weighted versions

$$C_{ij} = \langle \delta \mathbf{r}_i(t) \delta \mathbf{r}_j^*(t) \rangle \qquad C_{ij}^w = \langle \mathbf{u}_i(t) \mathbf{u}_j^*(t) \rangle \qquad \mathbf{C} = \mathbf{M}^{-1/2} \mathbf{C}^w \mathbf{M}^{-1/2*} \qquad (5)$$

being $\langle \rangle$ the time average (the vectors product is external)[a]. Both **C** and $\mathbf{C}^w$ are Hermitian, and therefore they admit orthonormal bases of eigenvectors. Considering first $\mathbf{C}^w$, in the harmonic case, using (3) one has

$$\mathbf{C}^w = \sum_{I,J} \mathbf{e}_I \mathbf{e}_J^* Q_I Q_J^* \langle \exp(-it(\omega_I - \omega_J)) \rangle_t = \sum_K Q_K^2 \mathbf{e}_K \mathbf{e}_K^* \;\Rightarrow\; C_{IJ}^w = Q_I^2 \delta_{IJ} \qquad (6)$$

implying that the normal modes $\mathbf{e}_I$ are also eigenvectors for $\mathbf{C}^w$, whose eigenvalues are the squared modes amplitudes $Q_I^2$. In other words, within the harmonic approximation the principal modes obtained from the mass weighted $\mathbf{C}^w$ coincide with the normal modes, and both **D** and $\mathbf{C}^w$ are diagonal on the $\{\mathbf{e}_I\}$ basis. However, the two operators bring different physical information, namely the frequencies (related to elastic constant and therefore "mechanical" information) or the amplitudes (related to the actual motion, and therefore a "dynamical" information), respectively. The connection between the two representations is possible when additional hypotheses are available. For instance, assuming the thermal equilibrium, the equipartition theorem implies that

$$Q_I^2 \omega_I^2 = kT \qquad (7)$$

The other way round, eqn (7) can be used after NMA and PCA are performed, to measure deviations from equilibrium.

PCA is widely used especially in the analysis of the motion of complex systems (e.g. biomolecules[Amadei (1993)]) whose dynamics is seldom at equilibrium, and focus on the determination of the main directions of the system motions, rather than on their time dependent behavior, which are averaged out in (5). In fact, when anharmonicity is present, $\mathbf{C}^w$ diagonalization still returns an orthonormal basis $\{\varepsilon_I\}$ and eigenvalues $X_I^2$, which, however do not generally coincide with eigenmodes and their amplitudes; therefore the motion projection onto $\varepsilon_I$ might include more than a single frequency. Furthermore, eqn (5) requires $\langle \rangle$ being an "average", not necessarily a "time average". Therefore it is sometime used with statistical ensembles, rather than real trajectories, possibly including thermodynamic information but loosing the dynamical one. In these conditions, clearly, the use of the mass weighted $\mathbf{C}^w$ – useful when the connection of NMA is feasible – appears less relevant. In this case, the non-mass weighted **C** [Daidone (2012)] gives a more direct physical interpretation in terms of displacements. Since also **C** is Hermitian, it admits orthonormal eigenvectors whose eigenvalues are true displacement amplitudes (i.e. not mass weighted amplitudes).

---

[a] Here, the complex conjugate of the displacement is included for generality, although all the formalism could also be carried on also considering a completely real case. (The * sign indicate the adjoint operation when applied to a vector or matrix, i.e. transposition and complex conjugation).

$$\mathbf{C} = \sum_K A_K^2 \eta_K \eta_K^* \quad \Rightarrow \quad C_{IJ} = A_I^2 \delta_{IJ} \tag{8}$$

{$\boldsymbol{\eta}_I$} represent the main directions of the system displacement in the given structures set and the projection $A_I(t)=\delta\mathbf{r}(t)\cdot\boldsymbol{\eta}_I$ already has the meaning of "mode coordinate". We observe however, that out of harmonic approximation, $\mathbf{C}^w$ and $\mathbf{C}$ explore different aspects of anharmonicity, especially in the case of hetero-atomic systems, due to the different weights of coordinates. Therefore it is worth evaluating both sets of eigenvectors in any case. {$\boldsymbol{\eta}_I$} generally differ from {$\boldsymbol{\varepsilon}_I$} and from {$\mathbf{e}_I$}, and their relationship are rather complex. Simple relationships hold only in the harmonic and homo-atomic case[b], when

$$C_{IJ}^w = mC_{IJ} \qquad \boldsymbol{\eta}_I = \boldsymbol{\varepsilon}_I = \mathbf{e}_I \qquad Q_I^2 = X_I^2 = mA_I^2 = \frac{kT}{\omega_I^2} \tag{9}$$

last equality holding only in case of thermal equilibrium. A summary of definitions and relationships is reported in Table 1.

Table 1: Summary of the notations used in the different approaches to analyze the collective vibrational excitations, their eigenvalues and eigenvectors and inter relations. (y) or (n) in the "Eigenv" column indicate wether the basis of eigenvector can be orthonormal.

|  | Input | Output | | | general | harmonic | homo-atomic |
|---|---|---|---|---|---|---|---|
|  |  | Eigenv | freq | amp |  |  |  |
| NMA | $\mathbf{D}$ | {$\mathbf{e}_I$}(y) {$\boldsymbol{\xi}_I$}(n) | $\omega_I^2$ |  | $\mathbf{u}_I(t) = Q_I(t)\mathbf{e}_I$ $\delta\mathbf{r}_I(t) = Q_I(t)\boldsymbol{\xi}_I$ $\boldsymbol{\xi}_I = \mathbf{M}^{-1/2}\mathbf{e}_I$ | $\mathbf{u}_I(t) = Q_I^0 \mathbf{e}_I\, e^{-i\omega_I t}$ $\delta\mathbf{r}_I(t) = Q_I^0 \boldsymbol{\xi}_I\, e^{-i\omega_I t}$ | $\boldsymbol{\xi}_I = \mathbf{e}_I/\sqrt{m}$ |
| PCAw | $\mathbf{C}^w$ | {$\boldsymbol{\varepsilon}_I$}(y) |  | $X_I^2$ | $\mathbf{C}^w = \mathbf{M}^{1/2}\mathbf{C}\mathbf{M}^{1/2*}$ | $\boldsymbol{\varepsilon}_I = \mathbf{e}_I$ $X_I^2 = Q_I^2$ | $\mathbf{C}^w = m\mathbf{C}$ |
| PCA | $\mathbf{C}$ | {$\boldsymbol{\eta}_I$}(y) |  | $A_I^2$ | $\dfrac{(\boldsymbol{\varepsilon}_J^* \mathbf{M}^{1/2}\boldsymbol{\eta}_I)}{(\boldsymbol{\varepsilon}_J^* \mathbf{M}^{-1/2}\boldsymbol{\eta}_I)} = \dfrac{X_J^2}{A_I^2}$ |  | $X_I^2 = mA_I^2$ $\boldsymbol{\eta}_I = \boldsymbol{\varepsilon}_I$ |
| SA vel | $\mathbf{v}, c_v$ |  | $\omega_I$ | $P(\omega_I)$ | $P(\omega) = \omega^2 F(\omega)$ | $P(\omega_I) = \omega_I^2 {Q_I^0}^2$ |  |
| SA dis | $\mathbf{u}, c_u$ | $\bar{\mathbf{u}}(\omega_I)$ | $\omega_I$ | $F(\omega_I)$ | $F(\omega) = |\bar{\mathbf{u}}(\omega_I)|^2$ | $F(\omega_I) = {Q_I^0}^2$ $\bar{\mathbf{u}}(\omega_I) = Q_I^0\, \mathbf{e}_I$ |  |

## 2.3. *Spectral analysis*

PCA is aimed at extracting the collective mode direction and amplitude from the trajectory, regardless the time dependent behavior. i.e. finding "space coherency". Conversely, the focus of spectral analysis (SA) is on the detection of characteristic frequencies of the system ("time coherency"). To this aim, the mass weighted displacement and velocity self correlation function are defined

$$c_u(t) = \frac{\left\langle \sum_i \mathbf{u}_i(t+t')\cdot\mathbf{u}_i(t') \right\rangle_{t'}}{\left\langle \sum_i \mathbf{u}_i(t')\cdot\mathbf{u}_i(t') \right\rangle_{t'}} \qquad c_v(t) = \frac{\left\langle \sum_i \mathbf{v}_i(t+t')\cdot\mathbf{v}_i(t') \right\rangle_{t'}}{\left\langle \sum_i \mathbf{v}_i(t')\cdot\mathbf{v}_i(t') \right\rangle_{t'}} \tag{10}$$

---

[b] Here non-degeneracy between frequencies is assumed. In case of degeneracy, a there is freedom of choosing different eigenvectors in the degenerate sub-space.

being $\mathbf{v} = \dot{\mathbf{u}}$. As previously, mass weighted and normal quantities can be considered, and give different results in case of hetero-atomic systems, analogous to those discussed above. Even in this case, a direct relationship with NMA in the harmonic approximation is obtained for mass weighted quantities, which are therefore here considered as primary observables. Similarly to PCA, this approach is based on analysis of trajectories, therefore it includes dynamical information on the actual behavior of the system. But at variance with it, the information on space correlations between specific atomic displacements is averaged out in the sum over *i*, while the information of time correlations is explicitly maintained. This also implies that in this case the average < > cannot be substituted by a statistical average. Characteristic frequencies are extracted by Fourier Transform (FT) [c]

$$F(\omega) = \frac{1}{2\pi} \int_{-\infty}^{+\infty} c_u(t) e^{-i\omega t} dt \qquad P(\omega) = \frac{1}{2\pi} \int_{-\infty}^{+\infty} c_v(t) e^{-i\omega t} dt \qquad (11)$$

Using eq (3) in (11) (possible within the harmonic approximation) it can be to show that

$$F(\omega) \propto \sum_I Q_I^{0\,2} \delta(\omega - \omega_I) \qquad P(\omega) \propto \sum_I \omega_I^2 Q_I^{0\,2} \delta(\omega - \omega_I) \qquad (12)$$

that fulfill the relationship $P(\omega)=\omega^2 F(\omega)$. Therefore $F(\omega)$ and $P(\omega)$ bear similar information [Thomas *et al* (2013)]: they display sharp peaks at the vibrational eigenfrequencies $\omega_I$. In the intensity spectrum $F(\omega)$, the peaks height is proportional to the squared (mass weighted) amplitudes of the mode, while in the power spectrum $P(\omega)$ lower frequencies are quenched, resulting in an equal eight peaks (if not degenerate) at thermal equilibrium (see eqn 7). Degeneracy, of course, proportionally increases the intensity of each peak. In the case of a crystal the vibrational states are continuously indexed by $\omega(\mathbf{k})$, therefore one can write at equilibrium $P(\omega) \propto g(\omega)$ being the latter the vibrational density of states VDOS[Han *et al* (2013)].

In the evaluation of velocity or displacement correlation functions, the information about single atom displacement is averaged out in (10). However, the possibility of evaluating the vibrations eigenvectors can be recovered directly from the FT of the displacements. Using Eq. (3) one has

$$\bar{\mathbf{u}}(\omega_I) = Q_I^0 \mathbf{e}_I \qquad \delta \bar{\mathbf{r}}(\omega_I) = \mathbf{M}^{-1/2} \mathbf{e}_I Q_I^0 \qquad (13)$$

(the over-bar indicates the FT operation). Therefore, the eigenvectors of vibrations are the FT (mass weighted) displacements evaluated at the mode frequency. In fact, the complete $\bar{\mathbf{u}}(\omega)$ vector would already include all the spectral information, since from eq (10) one has

$$F(\omega) \propto |\bar{\mathbf{u}}(\omega)|^2 \qquad P(\omega) \propto |\bar{\mathbf{v}}(\omega)|^2 = \omega^2 |\bar{\mathbf{u}}(\omega)|^2 \qquad (14)$$

where | | indicates norm of the whole coordinate (3N-sized) vector. However, the evaluation of *F* and *P* from (10) and (11) is faster (and more accurate, see below) when only the spectral properties (not the eigenstates) are needed.

Anharmonicity produces a broadening of the peaks and their shift. However the concepts of SA can still be used: the eigenfrequencies can be defined by the peaks location, while the peak width is related to the inverse vibration coherency time; the vibration eigenvectors can still be approximately described by Eq. (13), evaluated at average peak frequency. In addition, different approaches can be combined: for instance, one could use PCA to find the eigenvectors, and subsequently perform the SA on the projected trajectories to obtain the $\omega_l$ and spectra of a single principal mode. On the other way round, one could perform the PCA of a single normal mode trajectory. Again, it should be kept in mind that different approaches give the same eigenvectors and eigenvalues in the harmonic case, but might reveal different aspect of anharmonicity, when present.

---

[c] *P* and *F* are real and positive, implying that *c*s are even (besides real, by definition).

### 3. Practical issues and examples

We implemented SA and PCA in a home-made software (available upon request). In the following we report the main practical issues. We illustrate applications using as test case a microcanonical *ensemble* simulation of molecular dynamics of $C_{60}$ and of a graphene sheet excited with specific modes. Namely, we excited the breathing mode on fullerene (see Fig 1) and a flexural phonon on graphene. We used the Tersoff-Lindsay [Tersoff (1988), Lindsay and Broido (2010)] potential to represent the interaction, which was previously tested in a number of similar systems [Camiola (2015)].

### 3.1. *Numerical estimators of the time correlation functions*

In the SA, the main issue is that the averages over *t'* must be evaluated over a finite and discrete trajectory obtained from a simulation [Koukaras et al (2015)]. This has two consequences: first the numerical estimators for the *c*s will be themselves evaluated on discrete times. Second, the finiteness of the trajectory implies the possibility of defining several estimators differing in the extremes of the sum at the numerator and/or denominator of Eq. (10). One can identify three types of estimators, labeled by $\Delta$, *max* and *bias*, all satisfying the condition $c(0)=1$

$$c^{bias}(t_i) = \frac{\sum_{j=1}^{N}\sum_{t_k=0}^{t_{max}-t_i} \mathbf{v}_j(t_i+t_k)\cdot\mathbf{v}_j(t_k)}{N_{t_{max}}\sum_{j=1}^{N}\langle\mathbf{v}_j^2\rangle} \qquad c^{max}(t_i) = \frac{\sum_{j=1}^{N}\sum_{t_k=0}^{t_{max}-t_i} \mathbf{v}_j(t_i+t_k)\cdot\mathbf{v}_j(t_k)}{N_{t_{max}-t_i}\sum_{j=1}^{N}\langle\mathbf{v}_j^2\rangle} \qquad c^{\Delta}(t_i) = \frac{\sum_{j=1}^{N}\sum_{t_i=0}^{\Delta} \mathbf{v}_j(t_i+t_k)\cdot\mathbf{v}_j(t_k)}{N_{\Delta}\sum_{j=1}^{N}\langle\mathbf{v}_j^2\rangle} \quad (15)$$

(here defined for the velocity correlation function but analogously defined for $c_u$, and for their mass weighted cases); $N_x$ is the number of discrete time points included in the summation interval indicated by *x*, and <> indicates the time average.

We first observe that $c_{bias}$ has a number of summation time steps less than $N_{tmax}$, and decreasing as $t_i$ increases. This implies that as $t_i$ increases, *c* value decrease also due to this unbalancing, which artificially superimpose to the physical behavior of *c*. In other words, the correlation function is multiplied by a linear decreasing (i.e. "triangular") cut function $f_T(t)=-|(t-t_{max})/t_{max}|$. This estimator, also called "biased" [Koukaras et al (2015)], identically vanishes at $t=t_{max}$ and its discontinuous derivatives in $t=0$ and $t=T$ brings an unphysical oscillating behavior around the spectral peaks, therefore it is generally not advisable. It is however numerically convenient because robust and parameterless.

The other two estimators are unbiased. i.e. in both cases the $N_x$ is chosen to correct the bias. Because the numerator time intervals are smaller than the maximum one $[0,t_{max}]$, there is some arbitrariness in the choice of the time interval for the evaluation of the time average at denominators. We assumed here the average evaluated over the whole time interval, which makes the estimator $c^{max}$ identical with the "unbiased" defined in ref. [Sathyabnaryana (2004)]. In this way, the average value at the denominator is always evaluated with the maximum numerical precision. The numerator of $c^{max}$, however, is evaluated with decreasing precision at increasing $t$, implying worsening of signal to noise ratio as $t$ approaches $t_{max}$. Conversely, $c^{\Delta}(t_i)$ has the same level of noise at all values of $t_i$. This is paid with a smaller definition interval of $c^{\Delta}$, namely $[0,t_{max}-\Delta]$. The choice of $\Delta$ value is demanded to the user, based on a compromise between good signal to noise ratio (requiring large $\Delta$) and large definition interval (requiring small $\Delta$). A good choice, used here, is $\Delta = t_{max}/2$. The $c^{max}$ and $c^{bias}$ estimators are reported in Fig 2 (a) for a 100ps run of the breathing mode of $C_{60}$ in black and cyan respectively. The correlation functions show clearly the presence of a dominant fast oscillation with period ~0.1 ps, which on the 100ps time scale is still present though enveloped with more slowly oscillating function (black lines). However at the end of the time range the correlation functions are still very different from zero, because the breathing mode is much more persistent than 100 ps. The biased estimator (cyan) forces the correlation function to 0 at 100ps. The unbiased estimator show a similar behavior, but its definition range is one half of the latter. The velocity and displacement based estimators are very similar (different versions are shown only for the velocity correlation function estimators).

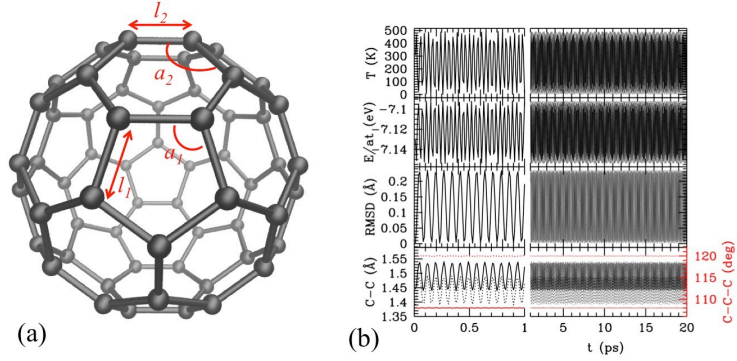

(a) (b)

Fig 1: Test case: $C_{60}$ fullerene (a) in which the breathing mode (= uniform expansion/contraction of the structure) is excited. In panel (b) are reported from top to bottom the kinetic energy (in equivalent temperature) the energy per atom, the root mean squared deviation from the starting structure, and the monitoring of the bond lengths (black) and angles (red) shown in panel (a). Because the excited mode is a uniform expansion, the angles vibration is negligible with respect to the bond-length oscillation.

### 3.2. *Evaluation of the power and intensity spectra*

Vibrational spectra are obtained by FT of the $c$s as from (11). Again, the FT integral must be numerically estimated due to the discretization and finiteness of the time coordinate. Due to the parity of $c(t)$ with respect to time reversal, the effective definition interval of $c$ is $[-T,T]$, $T$ being $t_{max}$ or $t_{max}-\Delta$ depending on the chosen estimator. Considering $c^{\Delta}$, the particular choice $\Delta=t_{max}/2$ leads to a definition interval as large as $t_{max}$, i.e. equal to the starting trajectory length. The definition interval in time domain determines the maximum discretization density in the frequency domain, namely $\delta\omega=\pi/T$ which is adopted as default in this work. We observe that using different $\Delta$s or using $c^{max}$ leads to an extension of time definition interval of $c$ (up to a maximum value of $2t_{max}$) and consequent an increase of points in the frequency domain. It might seem that this is an increase of information content of the spectrum, evaluated at doubled sampling frequency. However, this is paid with an increase in the noise in $c$ (and in the spectrum): as said, they are evaluated on a larger time interval (or with a larger sampling frequency) but less accurately. The lost in accuracy is uniform in time in $c^\Delta$ (and larger $\Delta$ is smaller) or concentrated at the time interval extrema in $c^{max}$. The choice $\Delta=t_{max}/2$ (leading to a frequency sampling density $\delta\omega=2\pi/t_{max}$) turns out a good compromise, with the additional advantage that times and frequency intervals are the same as the trajectory $\mathbf{u}(t)$, which allows a direct comparison of the eigenvalues from trajectory (see next section).

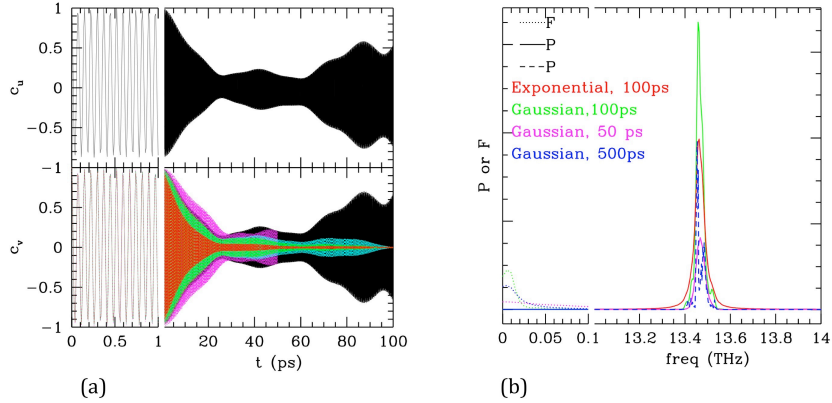

Fig 2: (a) correlation functions of the displacements (top) and velocities (bottom) in a 100 ps simulation of the "breathing" model of $C_{60}$. Velocity correlation function estimators are shown in all versions: Black = $c^{max}$; magenta= $c^\Delta$ ($\Delta=t_{max}/2$); cyan= $c^{bias}$; green = $c^{max}$ with Gaussian cut; red = $c^{max}$ with exponential (Lorenz) cut. In Gaussian cut the parameter $\alpha$ is 1/3, in the exponential cut it is 1/4.5. In the two plots, the first ps is expanded to show the fast oscillation, which is squeezed and becomes invisible in the normal view. (b) Intensity and power spectra obtained using the Gaussian cut in different lengths runs (green) and exponential (red) cut. The blue dashed line is the same as the green one, but with a 500ps long run. The magenta line is the is obtained from the Gaussian cut of $c^\Delta$. Power and intensity spectra are both shown, but basically indistinguishable at high frequencies. An expansion of the low frequency section show the intensity spectra as dotted lines.

On the other way round, the discretization interval in time determines the frequency interval definition, therefore $\omega_{max}=2\pi/\Delta t$. Being the FT periodically defined outside this interval, the minimal definition interval is often chosen among one of the two $[0,2\pi/\Delta t]$ or $[-\pi/\Delta t,\pi/\Delta t]$. In addition, it should be bear in mind that $c(t)$ is real and symmetric, therefore its FT is real and symmetric. Therefore in conclusion, the appropriate frequency interval for the spectra definition is $[0,\pi/\Delta t]$.

The finiteness of the time range definition brings a specific issue during the FT: the discontinuity at the boundaries of the time interval produces unphysical oscillations in the FT. Therefore, $c(t)$ should vanish at $T$, possibly "regularly" i.e. also its derivative should vanish. Usually, in real systems with large amount of anharmonicity, $c(t)$ displays a spontaneous decay physically related to decoherence and finite life time of modes. This solves the problem because manifests as vanishing of the $c(t)$ within the lifetime of the mode. However, if simulations are not long enough to observe the physical decay, or if the system has particularly persistent modes, an artificial decay envelope function must be used. Here we considered the following:

$$c(t_i) \to c(t_i)f_T(t_i) \qquad f_T = f_T^G = e^{-\frac{t^2}{2(T\alpha)^2}} \qquad f_T = f_T^L = e^{-\frac{t}{T\alpha}} \qquad f_T = f_T^{bias} = -\frac{|t-t_{max}|}{t_{max}} \qquad (16)$$

The different cut function brings different convolution function in FT, i.e. different line shapes: Gaussian for the Gaussian cut, Lorentzian for the exponential cut and oscillating for the linear cut ($\propto \sin^2(\Delta\omega)/\Delta\omega^2$). Fig 2(a) reports the correlation functions of the fullerene breathing mode simulations with the Gaussian cut (green) and exponential cut (red), compared with the linear cut produced by the biased estimator. As it can be seen, in spite of the cut, the slow oscillation enveloping the fast dominant one are still visible. However, in order to obtain a similar depression of the correlation function at $t_{max}$, a smaller value of $\alpha$ (i.e. shorter cut) must be used in the exponential form with respect to the Gaussian one. Therefore, the envelope shape remains more visible in the Gaussian cut, which is generally more advisable to detect the slower frequencies. On the other hand, the Lorentzian line shape is often considered "more physical".

Fig 2 b shows the comparison of spectra obtained from velocities (power spectrum P) or displacements (intensity spectrum F) with different cuts. The exponential cut shows a single Lorentzian peak at the dominant frequency (at 13.5THz), while the Gaussian cut shows a more structured peak, revealing at east three secondary peaks, responsible of the beats clearly visible in the envelope of correlation functions. Clearly, more defined spectra can be obtained with longer runs: the blue dashed line is the power spectrum obtained with a 500 ps run. The band splits in at least two main peaks and a number of secondary peaks. These are less symmetric deformation modes at similar frequency than the breathing mode, strongly correlated with it. Once coherently renormalized, F and P at those frequencies are basically superimposable. However at ~0.01THz in the F spectra a slow mode appears. At those low frequencies there are no internal modes. However, due to accumulation of unavoidable integration errors in such a long run and to the spherical symmetry of the system, roto-translational modes tends to excite, which are suppressed in the power spectrum.

### 3.3. *Eigenvectors of vibration: comparison between the PCA and the SA*

PCA directly outputs the eigenvectors of principal modes, but it is also possible to extract the eigenmodes of vibration at a given frequency directly from the trajectory, and to project the trajectory onto them to separate single modes. According to eqn (3), the eigenvector of vibration is the FT of the trajectory evaluated at the given frequency. However, as in the case of correlation functions, the trajectory does not vanish at the extreme of integration. Therefore one must rather evaluated the FT of an estimator, smoothly cutting trajectories:

$$\bar{\mathbf{u}}(\omega) = \sum_{0<t_i<t_{max}} \mathbf{u}(t_i) g_{t_{max}}(t_i) e^{-i\omega t_i} \qquad g_{t_{max}}(t_i) = e^{-\frac{(t_i - t_{max}/2)^2}{2(\beta t_{max})^2}} \qquad (17)$$

Then, ideally, according to (3) and (4), single modes projections and eigenvectors can be obtained evaluating (3) at a single frequency $\omega_I$. However, in the real case, peaks have a finite width, therefore the projected trajectory is appropriately estimated by the sum

$$\mathbf{u}_I(t) = \frac{1}{g_{t_{max}}(t)} \sum_{\omega_I - \delta\omega}^{\omega_I + \delta\omega} \Re e[\bar{\mathbf{u}}(\omega) e^{i\omega t}] \qquad \mathbf{r}_I(t) = (\mathbf{r}_{0,1} + \delta\mathbf{u}_1/\sqrt{m_1}, \cdots, \mathbf{r}_{0,n} + \delta\mathbf{u}_n/\sqrt{m_n}) \qquad (18)$$

being $[\omega_I-\delta\omega, \omega_I+\delta\omega]$ an interval around the mode peak. The vibration eigenvector could be estimated from the value assumed by the displacement at the time of the maximum elongation. Again, due to the finite width of the frequency interval, this definition can be approximated by different estimators. Two possible are

$$Q_I^0 \mathbf{e}_I = \bar{\mathbf{u}}(\omega)\big|_{\omega_I} \quad \text{and} \quad Q_I^0 \mathbf{e}_I = \sum_{\omega_I-\Delta\omega}^{\omega_I+\Delta\omega} \bar{\mathbf{u}}(\omega) \tag{19}$$

namely the elongation evaluated on the peak value or in a small interval around it. We observe that $\bar{\mathbf{u}}(\omega)$ is complex. Its phase, when evaluated at a single frequency, reflects the possibility of shifting the time origin, and is irrelevant. Therefore the first of estimators in (19) could be evaluated considering the norm of $\bar{\mathbf{u}}(\omega)$. Conversely, in the second estimator the relative phases at different frequencies cannot be neglected. The peak frequency and frequency interval is to be evaluated by means of a previous calculation of $P(\omega)$ or $F(\omega)$.

Fig 3 reports a comparison of the PCA and SA evaluated on the trajectory of Fig 1. The PCA shows three main modes plus other 3-4 minor ones. Similarly the SA (inset) shows a peak and 3-4 shoulders. The eigenvalue corresponding to the first principal mode is represented in the structure with green arrows and is clearly the pure breathing mode. The result of (19) evaluated with the integral estimator is represented in the structure with red arrows and it is the superposition of several modes, nearly degenerate with the breathing one.

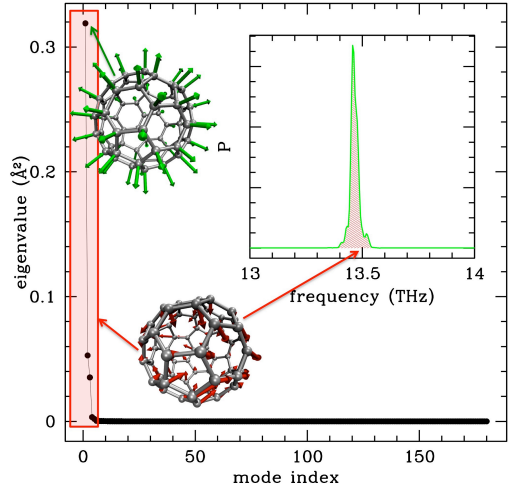

Fig 3: Illustration of the PCA and principal modes spectrum (black dots) and power spectrum (insect) evaluated on the same simulation (the one represented in Fig 1). The modes are represented with arrows starting from the atom in equilibrium configuration and whose length is proportional to the elongation in the corresponding mode. The structure with green arrow represent the first principal mode, while the figure with red arrows represent the mode located at ~13 THz in the power spectrum. It is a mixture of modes, which are those enclosed in the red rectangle in the PCA spectrum.

As a second example, we consider an extended periodic system namely the well know 2D crystal graphene [Geim and Novoselov (2017)]. Here we represent it by a previously well established scheme[Camiola (2015)] using an approximately squared simulation supercell of ~4.4 nm side (720 C atoms). We excited a stationary wave imposing a sinusoidal vertical displacement from the flat equilibrium configuration of amplitude ~1.2 Å and the same wavelength as the cell side. The system oscillates with a frequency of ~0.5 THz, according to the dispersion curve at this wavelength [Farchioni et al (2017)]. The analysis of velocity correlation function and of its spectrum Fig 4 (a and b), however, reveals the presence of a secondary peak at ~10 Thz. The analysis of eigenvectors according to (19) reveals the dominant mode at ~0.5THz is the out of plane sinus-like oscillation as expected (Fig 4b, structure with green arrows), while the fast oscillation is an in-plane longitudinal mode with wavelength ½ of the previous one. While a complete discussion about the coupling of these two modes is reported in a forthcoming paper[Farchioni *et al* (2017)], here we observe that this mode is due to the different sheet elongation occurring on the crests and on the nodal points of the flexural wave.

The PCA gives further insight (Fig 4). The first principal mode is the flexural one, and the second is the longitudinal one, as expected (see Fig 4a). However, the PCA reveals a third non negligible eigenvalue corresponding to a flexural mode with wavelength 1/3. In order to see if these principal modes also correspond to frequency eigenmodes, we evaluated the spectra of the corresponding projected trajectories (eqn 4. using $\varepsilon_I$ as projector vector, Fig 4 b, c and d). The projection on first principal mode (Fig 4b) gives a monochromatic spectrum, confirming that the first principal mode basically coincides with the first normal mode. The projection onto the second principal modes also acts nicely as a frequency filter, basically maintaining only the secondary peak at 10THz (the longitudinal mode), although a weak battement is visible, indicating that probably this mode is not pure. The third mode reveals more complex: it displays two peaks around 10THz, roughly corresponding to the frequencies of the flexural mode at those wavelength, but also a weaker peak at ~2THz. While the physical description of these features is beyond the scope of this work, we remark that this is the demonstration that principal modes may differ from normal modes, especially when anharmonicity comes into play. This particularly happens in flexural modes at short wavelengths, for which anharmonicity might appear at smaller values of the amplitude.

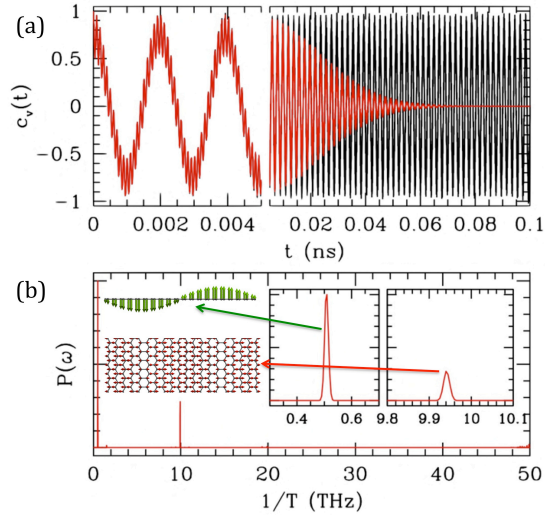

Fig 4**:** Spectral analysis and eigenvector evaluation of a sinusoidally excited graphene sheet. (a) velocity autocorrlation function: black line=bare, red line=gaussian cut (b) Power spectrum and its enlargement at the peak frequencies (insets). An arrows representation of the normal modes is reported within the plot.

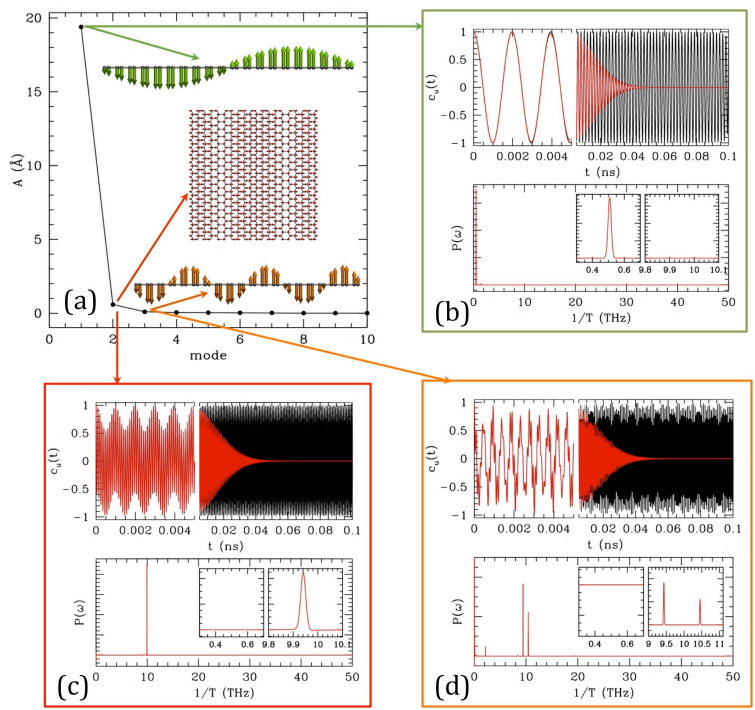

Fig 5**:** PCA (a) and spectral analysis of the projected trajectories on the first (b), second (c) and third (d) principal mode. The modes representation is reported in (a) within the plot representing the amplitude (squared root of the eigenvalue). (b), (c) and (d) reports displacement autocorrelation function (black lines= bare, red, Gaussian cut) and its spectra (enlarged in specific regions) for each mode projection.

## 4. Summary and conclusions

In summary, in this work we have reviewed and compared the main methodologies to extract vibrational information from a molecular dynamics simulation, namely the spectral analysis of time self-correlation functions and the principal component analysis of trajectories. The former focuses on the detection of time coherence and the latter on the detection of space coherence of the motion within a dynamical trajectory. As a consequence, the former returns characteristic frequencies and the latter modes and their amplitude of excitation. We showed that the two different points of view coincide in the harmonic case and practically demonstrate this on a couple of examples of outmost physical relevance, namely fullerene and graphene. We report issues related to the need of evaluating numerical estimators on finite and discrete trajectories and we give hints to solve them in the implementation. (Our own software – compatible with DL_POLY [Todorov et al (2006)] molecular dynamics simulation code and with VMD [Humphrey, et al(1996)] visualization code is available upon request[d]).

We also show what happens when harmonicity condition fails, and the concept of normal mode diverges from that of principal mode. The application in parallel or in cascade of the two methodologies reveals their different potentialities. PCA is usually more efficient in finding different modes especially when they are nearly degenerate in frequencies; however, the spectral analysis reveals that in general these modes are not pure frequency eigenstates. We show that, especially in case of anharmonicity and mode coupling, only the combination of the two methods can gives a physically insightful global view.

---

[d] Contact for information for software: dario.camiola@gmail.com

**Acknowledgments**

We thank Dr Vittorio Pellegrini for useful discussions. We gratefully acknowledge the financial support by the EU, 7th FP, Graphene Flagship (Contract No. NECT-ICT-604391) and Horizon2020, Graphene Core 1, (Grant Agreement No. 696656), and National Group of Mathematical Physics (GNFM-INDAM).